\documentclass[12pt]{article}
\usepackage{authblk}
\usepackage{geometry}
\usepackage{amsmath}
\usepackage{amssymb}
\usepackage{amsthm}
\usepackage{braket}
\usepackage{mathrsfs}
\usepackage{subfigure}
\usepackage{customcommands}
\usepackage{Nettastyle}

\title{A Holographic Argument for the Penrose Inequality in AdS}

\author[1,2]{Netta Engelhardt}
\author[3]{and Gary T. Horowitz}

\affiliation[1]{Department of Physics, Princeton University, Princeton, NJ 08544, USA}
\affiliation[2]{Gravity Initiative, Princeton University, Princeton NJ 08544, USA}
\affiliation[2]{Department of Physics, University of California Santa Barbara, CA 93106, USA}

\emailAdd{nengelhardt@princeton.edu}
\emailAdd{horowitz@ucsb.edu}

\abstract{We give a holographic argument in favor of the AdS Penrose inequality, which conjectures a lower bound on the total mass in terms of the area of apparent horizons. This inequality is often viewed as a  test of cosmic censorship. We further find a connection between the area law for apparent horizons and the Penrose inequality. Finally, we show that the argument also applies to solutions with charge, resulting in a charged Penrose inequality in AdS.	
}

\begin{document}

\maketitle

\section{Introduction and Summary}
\label{sec:intro}

Cosmic censorship, which states that regions of arbitrarily large spacetime curvature are invisible to asymptotic observers, is one of the oldest  conjectures about general relativity. It is also one of the most important: if it is true, general relativity is sufficient to predict everything that happens outside black holes, while its failure raises the possibility of directly observing astronomical effects of quantum gravity.  Despite its clear significance,  however, it remains unproven. In the absence of a proof, theoretical tests of cosmic censorship are of significant value.

In the early 1970's, Penrose~\cite{Pen73} proposed the following test of cosmic censorship: suppose one is given asymptotically flat initial data for general relativity with ADM mass $M$ and an apparent horizon  $\sigma$ with area $A[\sigma]$. Assuming cosmic censorship, $\sigma$ lies inside a black hole which is expected to settle down to a stationary Kerr solution.  Under evolution, the area of the event horizon cannot decrease and the total mass cannot increase. (Energy might be radiated away to null infinity, so the total (Bondi) mass may decrease.)  If the final black hole is described by the Schwarzschild solution, then $G M_{BH} = \sqrt {A_{BH} /16\pi}$. Since angular momentum decreases the horizon area, a final Kerr black hole satisfies $G M_{BH} \ge \sqrt {A_{BH} /16\pi}$. Since the initial quantities satisfy $M\ge M_{BH}$ and $A[\sigma]\le A_{BH}$, this gives an immediate prediction that all initial data must satisfy
\be\label{Penrose}
G M \ge \left(\frac{A[\sigma]}{16\pi} \right)^{1/2}. 
\ee
This is known as the Penrose inequality. It is a stronger form of the positive mass theorem, $M\ge 0$, and is conjectured to hold in the presence of an apparent horizon. A violation of this inequality would indicate a failure of cosmic censorship.\footnote{The converse, however is false: a proof of the Penrose inequality is \textit{not} tantamount to a proof of cosmic censorship. This is clear in more than four spacetime dimensions where the Penrose inequality might be true, but cosmic censorship fails.}

As stated, the inequality appears very difficult to prove.  Mathematicians have primarily focused on a Riemannian version of this inequality, which refers to an asymptotically flat Riemannian three-dimensional manifold with nonnegative scalar curvature and a minimal surface $\sigma$. This can be taken as initial data for a solution to Einstein's equation with zero extrinsic curvature and positive  energy density. In the resulting spacetime, the minimal surface is an apparent horizon.  After much effort, a complete proof of this Riemannian inequality was finally given in 2001, first for a single connected minimal surface~\cite{Huisken:2001} and then for several minimal surfaces \cite{Bray:2001}. Since not all asymptotically flat Lorentzian solutions to Einstein's equation necessarily admit such initial data, the general inequality remains open \cite{Mars:2009cj}.

A similar inequality has been conjectured for asymptotically anti-de Sitter (AdS) initial data with an apparent horizon $\sigma$~\cite{ItkOz11}.
The same arguments involving cosmic censorship and black holes settling down to a stationary solution lead to the conclusion that the mass $M$ and area $A[\sigma]$ of an asymptotically AdS initial data with apparent horizon $\sigma$ in 4D should satisfy\footnote{We are setting the AdS radius to one throughout this paper.} 
\be\label{AdSinequality}
GM \ge \left(\frac{A[\sigma]}{16\pi} \right)^{1/2} + \left(\frac{A[\sigma]}{16\pi} \right)^{3/2}.
\ee
Since  the horizon radius of a Schwarzschild AdS black hole, $r_+$, is related to its mass
 by $M = (r_+ +r_+^3)/2$, this is just the statement that $A[\sigma]$ is bounded from above by the horizon area of a static black hole with the same mass.
Despite some partial results (see, e.g., \cite{LeeNev13, BakSke14, Lima:2015,Husain:2017cmj}) this conjecture is largely open.

So far we have been assuming four-dimensional spacetimes, but there is no obstruction to considering the Penrose inequality in higher dimensions. The general form of the Penrose inequality in AdS in higher dimensions is given in~\cite{ItkOz11};  a proof of the Riemannian Penrose inequality for asymptotically flat Riemannian manifolds of dimension less than eight is given in \cite{Bray:2009}.  In higher dimensions, however, the Penrose inequality loses its connection with cosmic censorship, since there are unstable black holes in higher dimensions that develop singularities on their horizon when they pinch off, violating cosmic censorship. However, this type of naked singularity  is rather mild, in that its resolution in quantum gravity  is almost certainly to let the horizon bifurcate (or change its topology) with only of order a Planck energy  emitted in the process.  It is still possible that some relaxed version of cosmic censorship, which permits such mild singularities but forbids large-scale violations, remains valid. Such a reformulation of cosmic censorship could still  imply the Penrose inequality.

Another possibility is that the Penrose inequality may be false as a broad conjecture about general relativity but could be valid for theories of gravity coupled to low energy matter fields  that admit a UV completion within quantum gravity. This is a statement that can be tested within the framework of holographic quantum gravity~\cite{Mal97, Wit98a, GubKle98}. The classical limit of holography relates classical properties of gravity to properties of a dual quantum field theory (QFT) and one can use this dual description to try to derive new inequalities. 

We will show that a precise formulation of the (Lorentzian) AdS Penrose inequality follows from standard ideas in holography without assuming cosmic censorship. The basic idea is very simple. Given the initial data above, it is possible to construct a spacetime with two asymptotic boundaries and the same mass $M$ on each boundary such that the dual two-boundary QFT state has the property that  the  reduced density matrix of one boundary, $\rho_0$, has von Neumann entropy $S[\rho_0]= A[\sigma]/4G \hbar$ \cite{EngWal17b, Engelhardt:2018kcs}.  This entropy is clearly less than the maximum entropy of any density matrix with the same energy $M$. But the bulk dual to a maximum entropy state in a microcanonical ensemble is the static AdS black hole \cite{Marolf:2018ldl}. So 
\be \label{eq:introPenrose}
A[\sigma] = 4G  \hbar S[\rho_0] \le 4G \hbar \max_{\rm fix \ M}  S[\rho] = A_{BH}(M),
\ee
where $A_{BH}(M)$ is the area of a static AdS black hole with mass $M$. After solving the right hand side for $M$ we recover (\ref{AdSinequality}). As we discuss in the next section, we will require an extra condition on the apparent horizon which is generically satisfied. Since this constitutes what we believe is the first general argument for a Lorentzian Penrose inequality from first principles, it is possible that the correct general form of the inequality \eqref{eq:introPenrose} also requires this extra condition.~\footnote{We thank T. Jacobson for discussions on this.} 

The fact that the Penrose inequality follows so simply from holography raises the possibility that holography might imply a relaxed version of cosmic censorship as described above.  Another piece of evidence in favor of this possibility is the following. Since our construction involves a spacetime with a wormhole, the bulk theory must satisfy the weak gravity conjecture by the arguments in~\cite{Harlow:2015lma}. While the relevance of the weak gravity conjecture to Penrose's inequality may not a priori be clear, an intriguing connection has been discovered between cosmic censorship and the weak gravity conjecture: it was found in \cite{Horowitz:2016ezu,Crisford:2017zpi}  that cosmic censorship can be violated in AdS in a theory involving only a Maxwell field coupled to gravity. However, under inclusion of a charged scalar field with mass and charge satisfying the weak gravity conjecture, the Einstein-Maxwell counterexamples to cosmic censorship  
require fine tuning and are not generic \cite{Crisford:2017gsb,Horowitz:2019eum}. This triumvirate connection between cosmic censorship, the weak gravity conjecture, and the Penrose inequality is therefore  suggestive that some principle that rules out large violations of cosmic censorship may be generically satisfied in holography.

It is possible to construct a quantum generalization of the construction in \cite{EngWal17b, Engelhardt:2018kcs}, where $A[\sigma]$ is replaced by the generalized entropy $4G \hbar S_{gen}$~\cite{QuantumAH}. A natural question is then whether holography implies a quantum generalization of the Penrose inequality. Indeed, in order to prove Penrose's inequality in the semiclassical regime, we need to replace the null energy condition ($T_{ab}k^{a}k^{b}\geq 0$ for all null vectors $k^{a}$) with the so-called quantum focusing conjecture~\cite{BouFis15}. The final statement of the AdS Penrose inequality, however, is not particularly interesting: $4G \hbar S_{gen}$ differs from $A[\sigma]$ by a perturbative correction involving the entropy of quantum fields on the background classical spacetime. Since it is a perturbative correction, it can only make a difference in the case when the classical Penrose inequality is saturated. But in that case, it reduces to the well known statement that the entropy of a quantum field on a static black hole background is maximized by the Hartle-Hawking state. 

Saturation of the Penrose inequality is interesting in its own right, as the existing proofs of the Riemannian Penrose inequality for asymptotically flat initial data show that it is saturated only for the Schwarzschild solution.  So the Penrose inequality provides a rigidity result for Schwarzschild black holes. However, in the context of holography this is not the case: maximum entropy static black holes need not be unique. For example, at low energy in $AdS_5\times S^5$, localized ten dimensional black holes have more entropy than  Schwarzschild $AdS_5 \times S^5$, while the situation is reversed at high energy. Clearly, there is a particular energy at which these two different static black holes have the same entropy. Even without including the $S^5$ (or other compact extra dimensions) surprisingly little is known about the uniqueness of static AdS black holes. Even Schwarzschild AdS has not been shown to be unique. The best one has is a proof that there are no nearby static black holes \cite{Chrusciel:2017dpp}.

There are applications of the AdS Penrose inequality  to the area law for apparent horizons \cite{Hay93, AshKri02, AshKri03, BouEng15a, BouEng15b} and a proposed quasilocal mass formula \cite{BouNom18}. We will discuss these applications in Sec. 3, after deriving the inequality in the next section. There is also a generalization to charged black holes, which we describe in Sec. 3.

\section{Construction}

In this section, we will first review the requisite concepts for our argument. This includes a review of holographic entanglement entropy, our assumptions about apparent horizons, and of the dual to the area of apparent horizons. We will then present our argument for the Penrose inequality. 

\label{sec:construct}
\subsection{Background Review}

\paragraph{Assumptions and conventions:} we work in the large-$N$, strong coupling limit of gauge/gravity duality.  We will make all of the same assumptions as~\cite{EngWal17b, Engelhardt:2018kcs}, including the null energy condition. Since our construction relies in a large part on~\cite{EngWal17b, Engelhardt:2018kcs}, we will give a rough sketch of that construction and refer the reader to the original papers for technical details. Finally, we assume reflecting boundary conditions at the asymptotic boundary. In this section we will restrict to  time-independent boundary sources for our original QFT state (we follow a construction that results in a new QFT state, to which this assumption may not apply). This restriction will be lifted in Sec.~\ref{sec:gen}. 

We make use of the HRT prescription for holographic entanglement entropy~\cite{RyuTak06, HubRan07} (see~\cite{RanTak17} for a review):
\be
S_{vN}[\rho]= \frac{A[X_{\text{HRT}}]}{4G^{(D)}\hbar},
\ee 
where $\rho$ is the reduced density matrix on a single connected component $B_{1}$ of the asymptotic boundary, $G^{(D)}$ is the bulk Newton's constant in $D=10$ or 11-dimensions (depending on whether we are in string theory or M-theory), and $X_{\text{HRT}}$ is the minimal area spacelike codimension-two surface in the full (10 or 11-dimensional) bulk which is \textit{(i)} a stationary point of the area functional and \textit{(ii)} is homologous to $B_{1}$. The original prescription works for arbitrary subregions, but we  will only need to apply it to complete components of the asymptotic boundary. It is often the case that the surface $X_{\text{HRT}}$ wraps the internal dimensions, so that $X_{\text{HRT}}=X_{\text{HRT}}^{(d)}\times \mathcal{Y}_{D-d}$, where the full spacetime is given by $M_{d}\times \mathcal{Y}_{D-d}$, and $M_d$ is asymptotically AdS. In this case, we obtain
\be
S_{vN}[\rho]= \frac{A[X_{\text{HRT}}^{(d)}]}{4G^{(d)}\hbar},
\ee 
where $G^{(d)}$ is the $d$-dimensional Newton's constant. 

\begin{figure}
\begin{center}
\includegraphics[width=0.3\textwidth]{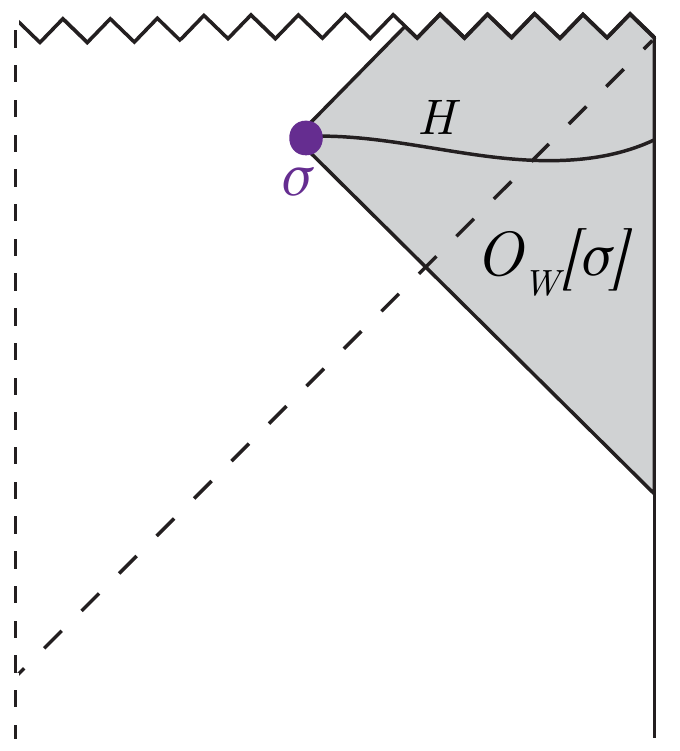}
\caption{Adapted from~\cite{Engelhardt:2018kcs}. An apparent horizon $\sigma$ (purple) and its outer wedge $O_{W}[\sigma]$ (shaded gray). By assumption there exists a Cauchy slice $H$ of $O_{W}[\sigma]$ on which $\sigma$ is the minimal area surface homologous to the boundary.}
\label{fig:minimar}
\end{center}
\end{figure}

We will also need a more recent addition to the holographic dictionary, which relates the area of a close variant of apparent horizons to a coarse-graining of the von Neumann entropy. Recall that an apparent horizon is a type of marginally trapped surface: that is, a compact, codimension-two surface $\sigma$ whose area is stationary under deformations in an outgoing null direction. Here \textit{outgoing} is defined with respect to the AdS boundary (in the situation that there are multiple connected components to the asymptotic boundary we define outgoing with respect to a particular connected component). More explicitly, if $k^a$ is the outgoing, future-directed orthogonal null vector to $\sigma$, and $h_{ab}$ is the induced metric on $\sigma$, then $\sigma$ is marginally trapped if the expansion
\be\label{expansion}
\theta \equiv h^{ab} \nabla_a k_b
\ee
vanishes everywhere on $\sigma$.
 The usual definition of an \textit{apparent horizon}  is the outermost, marginally trapped surface on a Cauchy slice $\Sigma$.  In \cite{EngWal17b, Engelhardt:2018kcs}, a closely related type of surface called a ``minimar'' surface was defined. A compact, marginally trapped surface $\sigma$ is said to be \textit{minimar} if: 
\begin{enumerate}
	\item  $\sigma$ is homologous to a (complete connected) component of the asymptotic boundary. That is, there exists a hypersurface $H$ such that $\partial H=\sigma \cup B$, where $B$ is a Cauchy slice of (a connected component of) the asymptotic boundary. The outer wedge of $\sigma$ -- the region spacelike to $\sigma$ and between it and the asymptotic boundary -- is the domain of dependence of $H$, and is denoted $O_{W}[\sigma]$ (see Fig.~\ref{fig:minimar}). 
	\item There exists a Cauchy slice $H$ of $O_{W}[\sigma]$ such that $\sigma$ is the minimal area surface on $H$ which is homologous to the boundary. 
	\item $\sigma$ is stable: consider the null geodesic congruence generated by $k^a$ with affine parameter $\lambda$ (with $\lambda = 0$ on $\sigma$) and let $\ell^{a}$ be an ingoing future-directed null vector orthogonal to surfaces of constant $\lambda$.
 Then there exists a parametrization of $\ell^{a}$ such that $k^{a}\nabla_{a}\theta_{(\ell)}\le 0$, where the expansion, $\theta_{(\ell)}$, is defined as in (\ref{expansion}) with $k_b$ replaced by $\ell_b$.  Equality can hold only if $\theta_{(\ell)}=0$ everywhere on $\sigma$.
 \end{enumerate}

Since apparent horizons are outermost on a Cauchy slice, they always satisfy requirement 1. One can show that generic apparent horizons satisfy the other two requirements also \cite{Engelhardt:2018kcs}. From here on, we will assume that our apparent horizons are minimar. 

\begin{figure}
\begin{centering}
\includegraphics[width=0.4\textwidth]{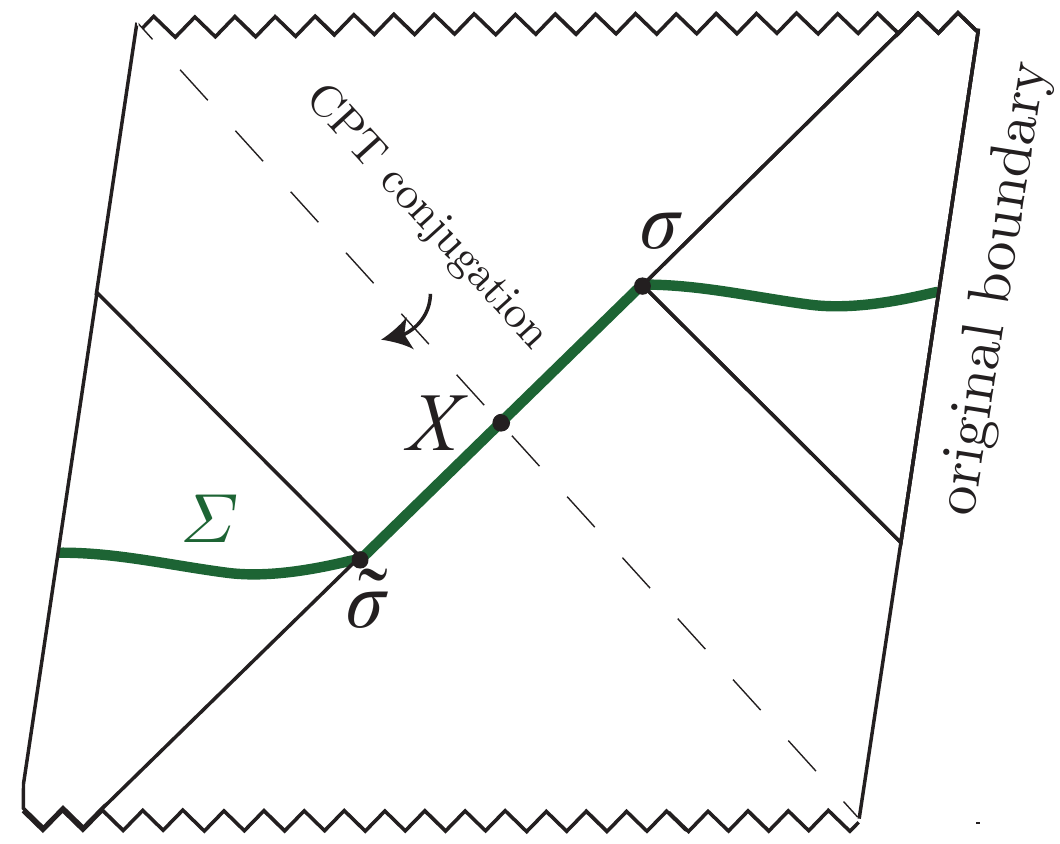}
\caption{Adapted from~\cite{Engelhardt:2018kcs}. The initial data slice $\Sigma$ (green) used to prepare the doubled spacetime with an HRT surface $X$ whose area is identical to the area of $\sigma$. Note that $O_{W}[\sigma]$ is fixed by taking the component of $\Sigma$ in $O_{W}[\sigma]$ to be identical in the doubled spacetime and in the original spacetime. The entanglement wedge bounded by $X$ and the right boundary is dual to $\rho_{0}$.}
\label{fig:DoubledSpacetime}
\end{centering}
\end{figure}

In~\cite{EngWal17b, Engelhardt:2018kcs}, it was argued that the area of apparent horizons is computed by a coarse-grained entropy called the outer entropy, obtained by maximizing the von Neumann entropy over all possible spacetimes that can be glued into the interior of $\sigma$:

\be
\frac{\mathrm{Area}[\sigma]}{4 G\hbar}= \max\limits_{\rho\in{\cal H}} S_{vN}[\rho]\equiv S^{\mathrm{outer}}[\sigma],
\ee
where ${\cal H}$ is the set of all QFT states with a semiclassical bulk dual which is identical to our original bulk in the region $O_{W}[\sigma]$. The proof identifies the state $\rho=\rho_{0}$ that maximizes $S_{vN}$ as above by explicitly constructing the dual bulk spacetime $(M',g')$: this spacetime has two asymptotic regions, agrees with the original spacetime $(M,g)$ on $O_{W}[\sigma]$, and has an HRT surface $X$ whose area is the same as the area of $\sigma$. The spacetime is constructed by preparing an initial data slice $\Sigma$ and time evolving it to generate the maximal evolution $M$.  It can then be shown that the HRT surface $X$ is spacelike separated from both asymptotic boundaries,  is null related to $\sigma$, and has the  same  area as $\sigma$ (see Fig.~\ref{fig:DoubledSpacetime}).\footnote{The regularity of the characteristic initial data specified on $\Sigma$ is expected to result in a locally unique Cauchy evolution~\cite{Ren90}; see e.g.~\cite{LukRod12,LukRod13}. The data satisfies the constraint equations, and is consistent with minimally-coupled scalar and Maxwell fields.}

Note that we are not assuming cosmic censorship, so it is possible that $M$ will have a Cauchy horizon and be extendible. As noted in~\cite{Engelhardt:2018kcs}, the same result still holds in this case. 

\subsection{An AdS Penrose Inequality}

The immediate conclusion that follows from the construction reviewed above is that there exists a spacetime $(M',g')$ in which the following equality holds:
\begin{equation}
A[\sigma] =A[X],
\end{equation}
where $X$ is the HRT surface of $(M',g')$. Assuming the HRT prescription, we have
\begin{equation}
A[\sigma] =A[X]=4 G^{(D)}\hbar S_{vN}[\rho_{0}],
\end{equation}
for some QFT state $\rho_{0}$ (and all quantities in the above equation are strictly finite in the large-$N$ limit, with the understanding that $G^{(D)} $ should be replaced by the appropriate power of 
$1/N$ according to the holographic dictionary)\footnote{We are taking the $N\rightarrow \infty$ limit in a way that keeps the ratio of the black hole radius to AdS length scale nonzero. If it does go to zero, one ends up with a black hole in aan symptotically flat spacetime and radiation can dominate the microcanonical ensemble.}. We note here two technical points. First, we are assuming that $(M',g')$ has a CFT dual. We have not proven that it always will, but we think that this is very likely.  Second, since the proof of the HRT proposal in~\cite{LewMal13, DonLew16}  applies to states that can be constructed by path integrals, and it is not obvious that $(M',g')$ can be constructed in this way, it is in principle possible that $(M',g')$ has a QFT dual but that its von Neumann entropy is \textit{not} computed by the area of the HRT surface. Thus our result could be framed as an exclusive alternative: either the Penrose inequality holds, or the HRT prescription is incomplete even in the regime of classical general relativity in the bulk. 

The outer wedge $O_{W}[\sigma]$ is by construction identical in $(M,g)$ and in $(M',g')$.  This immediately implies that asymptotic charges are identical in both spacetimes. In particular, the total mass within $O_{W}[\sigma]$ is identical in both spacetimes. This in turn implies that the QFT stress tensor integrated on any slice  of $\partial M\cap O_{W}[\sigma]$ -- the QFT energy $E$ -- is identical in $\rho$ and $\rho_{0}$ to leading order in $1/N$. Since we have assumed that any sources are time independent in $O_{W}[\sigma]$, this energy is independent of time.

Since $\rho_{0}$ is a state (on one connected component of the asymptotic boundary) with energy $E$, its entropy must be smaller than the entropy in the microcanonical ensemble:

\begin{equation}
G^{(D)}\hbar \, S_{vN}[\rho_{0}] \leq  G^{(D)}\hbar \max\limits_{E\pm \delta E} S_{vN}=G^{(D)}\hbar \, S_{vN}[\rho_{\text{micro}}] 
\end{equation}
where the right hand side is a maximization of $S_{vN}$ over all QFT states with energy in the range $E\pm \delta E$ at a fixed boundary Cauchy slice, where $\delta E $ is much larger than the difference between energy eigenvalues but much smaller than $E$. It has recently been argued in~\cite{Marolf:2018ldl} using the Euclidean path integral  that the bulk dual of the microcanonical ensemble is a static black hole of mass $E$, whose Bekenstein-Hawking entropy is precisely $S_{vN}[\rho_{\text{micro}}]$. 

We thus find:
\begin{equation}\label{AreaCompare}
A[\sigma] \leq  A_{BH}[E],
\end{equation}
Since we have not required  the dominating static black hole or $\sigma$ to be be a product with the internal space,  this formula applies to the full ten or eleven dimensional spacetime.   When the spacetime is asymptotically AdS$_{4}\times {\cal Y}^{7}$ and both $\sigma$ and the dominating saddle of the microcanonical ensemble are products with ${\cal Y}^{7}$, we recover precisely Eq.~\eqref{AdSinequality}, the Penrose inequality in four-dimensional  asymptotically AdS spacetimes.

\section{Generalizations and Applications}
\label{sec:gen}
\subsection{Time Dependent Sources}
In the previous section, we restricted to time-independent sources in the QFT to simplify the construction. We will now relax that condition and allow arbitrary time-dependent sources on the boundary. Since time-dependent sources will by definition result in changes to the total energy, we will need to be more precise about the asymptotic energy that goes into Eq.~\eqref{AreaCompare}. Operating under our prior assumption of reflecting boundary conditions, turning on boundary sources typically increases the energy, so if we choose to evaluate the  energy at a time slice $t=t_{1}$ on the boundary, it will generically be smaller than the  energy at a boundary time slice $t=t_{2}>t_{1}$. 

To obtain the tightest bound, we consider smooth spacelike cross-sections of the boundary that are contained in  $O_{W}[\sigma]$ and compute the energy on each. We then take the minimum of these energies, $E_{min}[\sigma]$.  We may execute the full construction above while keeping $E_{min}[\sigma]$ fixed, which yields the general inequality:

\be
A[\sigma]\leq A_{BH}[E_{min}[\sigma]]
\ee

\begin{figure}
\begin{centering}
\includegraphics[width=0.4\textwidth]{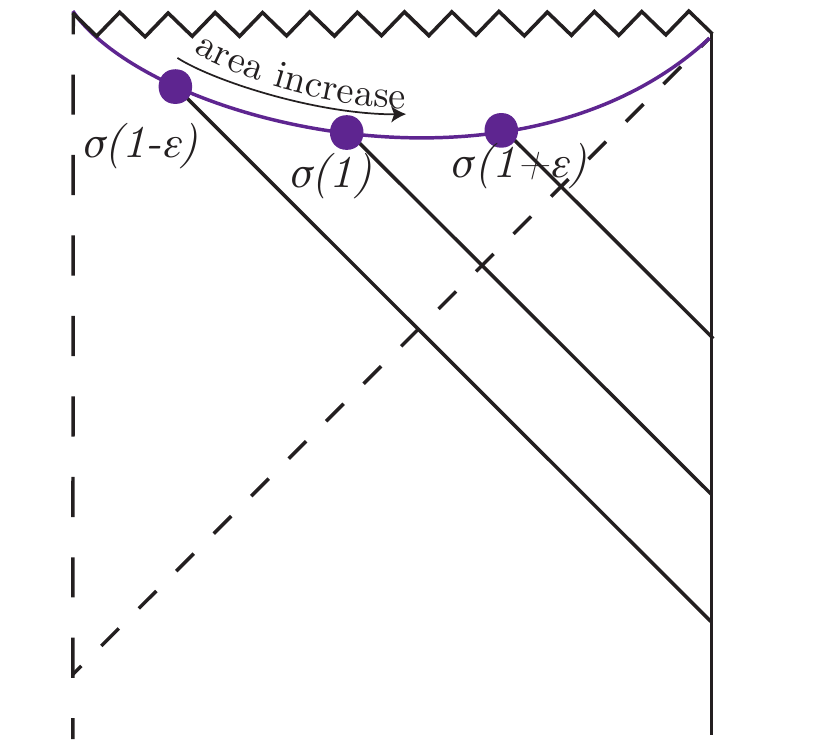}
\caption{The area increases in a spacelike direction
along a future holographic screen, which corresponds to time increase on the boundary. Allowing time-dependent sources on the boundary results in an increase in $E$. We maximize $S_{vN}$ subject to $E$ at the past boundary of $O_{W}[\sigma]$, so the mass in the Penrose inequality increases correspondingly with the apparent horizon area increase. }
\label{fig:timedep}
\end{centering}
\end{figure}

Recall now that hypersurfaces foliated by marginally trapped surfaces -- so-called future holographic 
screens~\cite{Bou99c}-- satisfy an area monotonicity theorem~\cite{Hay93, AshKri02, AshKri03, BouEng15a, BouEng15b}. In particular, the spacelike component of a future holographic screen is foliated by minimar surfaces. If the (minimar) apparent horizons in the foliations are labeled $\sigma(r)$, with $r$ the foliation parameter, then evolving forwards to increasing $r$ along the holographic screen corresponds to evolving along the boundary towards the future: the past boundary of $O_{W}[\sigma(1)]\cap \partial M$  is to the past of the past boundary of $O_{W}[\sigma(1+\epsilon)]\cap \partial M$. See Fig.~\ref{fig:timedep}. The energy increase due to boundary sources corresponds to an area increase, and the two are related via the Penrose inequality. 

In~\cite{EngWal17b, Engelhardt:2018kcs}, the area increase was interpreted as a thermodynamic second law: a coarse-grained QFT entropy increase. Here we see that this thermodynamic entropy increase is also related to an energy increase via the Penrose inequality. 

\subsection{Connection to Quasi-Local Mass}

There is an interesting connection between our derivation of the Penrose inequality and a recent definition of a quasi-local gravitational mass associated to a
``normal'' (non-trapped, $\theta_{k}>0$) surface.   Bousso et. al. \cite{BouNom18} proposed that the  outer entropy of a normal surface $\nu$ should be thought of as defining a quasi-local mass $M_{\nu}$ associated to $\nu$. 
They defined $M_\nu$ using the relation between mass and area of a Schwarzschild black hole. However, since we are considering asymptotically AdS spacetimes, it seems more appropriate to use the Schwarzschild AdS solution. In four dimensions, this AdS version of the proposal in \cite{BouNom18} is
\be\label{quasi}
2 G M_{\nu} = (a S^\mathrm{outer}[\nu])^{1/2}    + (a S^\mathrm{outer}[\nu])^{3/2}
\ee 
where  $a = G \hbar/\pi$.
Bousso et. al.~\cite{BouNom18} also construct a generalization of the doubled spacetime construction reviewed in Sec.~\ref{sec:construct} for normal surfaces. They provide initial data for a spacetime $(M'',g'')$ with an HRT surface $X_{\nu}$ whose area is given by   $S^{\mathrm{outer}}[\nu]$, and where the outer wedge $O_{W}[\nu]$ is unchanged. 

Now we can apply our above argument. Let $\rho_{1}$ be the dual to one side of $(M'',g'')$. The von Neumann entropy of $\rho_{1}$ is given by the area of $X_{\nu}$, and is smaller than the von Neumann entropy in the microcanonical ensemble with the same energy. Therefore:
\be
4 G\hbar\,  S^{\mathrm{outer}}[\nu] = A[X_{\nu}]= 4G\hbar \, S_{vN}[\rho_{1}]\leq 4 G\hbar \, S_{vN}[\rho_{\text{micro}} ]= A_{BH}(M)
\ee 
Since the relation between $M$ and $A_{BH}$ is similar to (\ref{quasi}), our argument shows that the quasi-local mass $M_{\nu}$ is bounded from above by the total mass $M$ (without assuming cosmic censorship). This is  
a desirable property of a quasilocal mass.

\subsection{Charged Black Holes}

There is a charged version of the Penrose inequality. We first consider the asymptotically flat case, and then generalize to asymptotically AdS. The same chain of arguments that led to  (\ref{Penrose}) shows that 
assuming cosmic censorship,  if one starts with initial data with mass $M$ and charge $Q$, and assumes that no charge can be radiated away, then the area of the initial apparent horizon must  satisfy
\be\label{QPenrose}
 \left(\frac{A[\sigma]}{16\pi} \right)^{1/2} \le \frac{1}{2}\left [GM + \sqrt{(GM)^2 - Q^2}\right ].
 \ee
 This is because if the final black hole is Reissner-Nordstrom, the inequality is saturated with $A = A_{BH}$ and $M = M_{BH}$. If the final black hole is rotating, the left hand side is reduced.  Referring back to the original quantities only reduces the left hand side further and increases  the right hand side. 
 Eq. (\ref{QPenrose}) can be viewed as a strengthening of the positive mass theorem in the presence of charge:  $GM \ge |Q|$.
 This argument also extends to AdS; the statement is imply  that the initial area cannot be greater than the area of a 
  Reissner-Nordstrom AdS black hole with the same mass and charge. 
 
 The holographic argument is  easily generalized to include charge. One can again construct a spacetime so that the dual state, $\rho_0$, has the same mass and charge as the original one and satisfies $A[\sigma] = 4G^{(D)} \hbar \, S_{vN}[\rho_0] $.  This is because the arguments in \cite{EngWal17b, Engelhardt:2018kcs} included a possible Maxwell field.  One can now maximize $S_{vN}[\rho]$ over all states holding the charge fixed as well as the mass.  The argument in \cite{Marolf:2018ldl} is easily generalized to include charge, with the result that the bulk dual to a maximum entropy state at fixed energy and charge is the maximum area static black hole with the same conserved quantities\footnote{We thank D. Marolf for a discussion about this.}.    This implies that $A[\sigma] < A_{BH}(M,Q) $ which is the charged Penrose inequality in AdS.
 As before, no assumption of cosmic censorship is needed in this derivation.
  
 It is worth noting that there is a slight difference in the treatment of electric and magnetic charges. Magnetic charges are simpler to incorporate in Euclidean path integrals for  two reasons. First,  the Maxwell field stays real after analytic continuation so one does not have to consider imaginary fields. Second, the standard gravitational path integral computes the partition function at fixed charge only for magnetic charge. For electric charge, the path integral computes the partition function at fixed potential. This is because the standard boundary condition keeps the potential $A_\mu$ fixed on the boundary, which is sufficient to compute magnetic charges but not electric ones. To compute the partition function at fixed electric charge, one needs to either modify the Maxwell action in the path integral, or  project the partition function onto a definite charge \cite{Hawking:1995ap}. Neither of these differences changes the final result.

\subsection*{Acknowledgments}
It is a pleasure to thank R. Bousso, X. Dong, R. Emparan, S. Fischetti, G. Gibbons, T. Jacobson, D. Marolf, and J. Santos for discussions. NE is supported by the Princeton University Gravity Initiative  and by NSF grant No. PHY-1620059. GH is supported in part by NSF grant PHY-1801805.
\bibliographystyle{JHEP}

\bibliography{all}

\end{document}